# Prediction of ultra-high ON/OFF ratio nanoelectromechanical switching from covalently-bound $C_{60}$ chains

Han Seul Kim,[a] Jhinhwan Lee,[b,c] and Yong-Hoon Kim[a,b,*]

[a] Graduate School of EEWS, KAIST, 291 Daehak-ro, Yuseong-gu, Daejeon 305-701, Korea.

[b] KI for the NanoCentury, KAIST, 291 Daehak-ro, Yuseong-gu, Daejeon 305-701, Korea.

[c] Department of Physics, KAIST, 291 Daehak-ro, Yuseong-gu, Daejeon 305-701, Korea.

## Abstract

Applying a first-principles computational approach, we have systematically analyzed the effects of [2+2] cycloaddition oligomerization of fullerene $C_{60}$ chains on their junction electronic and charge transport properties. For hypothetical infinite $C_{60}$ chains, we first establish that the polymerization can in principle increase conductance by several orders of magnitude due to the strong orbital hybridizations and band formation. On the other hand, our simulations of the constant-height scanning tunneling microscope (STM) configuration shows that, in agreement with the recent experimental conclusion, the junction electronic structure and device characteristics are virtually unaffected by the $C_{60}$ chain oligomerization. We further predict that the switching characteristics including even the ON/OFF-state assignment will sensitively depend on the substrate metal species due to the Fermi-level pinning at the substrate-side contact and the subsequent energy level bending toward the STM tip-side contact. We finally demonstrate that a force-feedbacked nanoelectromechanical approach in which both of the $C_{60}$–electrode distances are kept at short distances before and after switching operations can achieve a metal-independent and significantly improved switching performance due to the Fermi-level pinning in both contacts and the large intrinsic conductance switching capacity of the $C_{60}$

---

[*] Corresponding author. Tel: +82 423501717. E-mail address: y.h.kim@kaist.ac.kr (Y.-H. Kim)

chain oligomerization.

# 1. Introduction

Adopting functional bistable molecules as the active element of nanoscale switches represents an attractive route toward next-generation memory devices [1]. In the effort to progress the fundamental science and practical realization of molecular switches, a potentially feasible scheme would be utilizing the polymerization of $C_{60}$ fullerene molecules based on [2+2] cycloaddition [2-8]. Indeed, Nakaya *et al*. recently realized the scanning tunneling microscope (STM) tip pulse-induced oligomerization and deoligomerziation of $C_{60}$ chains in ultrathin $C_{60}$ films, demonstrating a possible ultra-high density memory device [9, 10]. After applying strong negative "writing" sample bias voltage pulses (~−3.5 V) to form linear oligomerized $C_{60}$ chains up to pentamers, they observed the local increase of tunnel resistance on individual $C_{60}$ molecule with a smaller "reading" sample bias (~+1.0 V). It was also shown that the bound $C_{60}$ molecules could be converted back to the individual unbound $C_{60}$ molecules by applying a large positive "erasing" sample bias (~+4.5 V). While using a single $C_{60}$ molecule as a data storage bit represents an impressive areal data density of 190 Tbit/in$^2$, the ON/OFF switching ratio achieved for the three-layer $C_{60}$ film probed at a constant-height STM mode was only 4 – 7.

In a previous study [11], using a first-principles approach [12-14], we investigated the electronic structures and charge transport properties of oligomerized $C_{60}$ chains symmetrically sandwiched between Au(111) and Al(111) electrodes and particularly emphasized the important role of the Fermi-level pinning that leads to the junction charge transport properties that are rather insensitive to electrode metal species. In view of the recent experimental realization of molecular switches based on the formation and breaking of $C_{60}$ chains [9, 10], we here extend our earlier work to theoretically compare the oligomerization-induced switching properties of the $C_{60}$ chain-electrodes system in two conditions: (1) the constant-height STM mode as realized in the experiment where the electrode-electrode distance is fixed during switching and (2) a force-feedbacked electrode mode where the distance between the two electrodes is self-adjusted to the length of the $C_{60}$ chain by the van der Waals forces as can be realized in a nanoelectromechanical device geometry [15] or a conducting-probe atomic force microscopy (C-AFM) [16]. The latter will be the basis of our proposal for a symmetric nanoelectromechanical device configuration that can achieve ultra-high ON/OFF switching ratios.

The finding in the STM experiments [9, 10] that particularly attracted our attention was that the change in the electronic properties of the $C_{60}$ chains does not play an important role for the observed switching phenomena. Because it is well-known that the STM signals originate from convoluted electronic structures of samples and STM tips or contacts, we start in this work by separating out the effect of the switching characteristics "intrinsic" to the $C_{60}$ chains from those of the tunneling gap size and elemental species of the electrodes. More specifically, we consider *infinite* $C_{60}$ chains and show that the strong orbital hybridizations upon [2+2] cycloaddition oligormerization should in principle lead to conductances of the bound $C_{60}$ chains several orders of magnitude higher (defined throughout this work as ON) than those of unbound counterparts (OFF). Then a relevant questions would be why the STM experiments showed the opposite switching-state assignment and small switching ratios and if there is a way to maximally exploit the intrinsically large conductance switching ratio of $C_{60}$ chains. To address these questions, we first scrutinize the STM junction configurations and find that the conductance switching ratio is dominated by the change in the tunneling gap distance between the top $C_{60}$ and the STM tip rather than the change of the intrinsic electronic properties of $C_{60}$ chains: Due to the efficient Fermi-level pinning of $C_{60}$ only to the substrate-side contact [11], the switching characteristics critically depend on the electrode metal species, which can completely change the direction of band bending or shifting along the $C_{60}$ chain. The strongly metal elemental dependent ON/OFF assignment (unbound = ON/bound = OFF for Au electrodes and the opposite for Al electrodes) emphatically demonstrates a critical role played by the electrode metal spe-cies in the important class of molecular electronic devices based on asymmetric electrode coupling. Finally, as an answer to the second question, we propose a mechanically-feedbacked nanoelectrochemical device architecture, and demonstrate that, due to the strong Fermi-level pinning character of $C_{60}$, it should realize the ultra-high ON/OFF switching ratios irrespective of electrode metal species.

## 2. Computational method

We applied a first-principles computational approach combining density-functional theory (DFT) and matrix Green's function (MGF) calculations. First, for the geometry optimizations of $C_{60}$ chain molecules and the junction models based on them, we performed DFT calculations using a modified version of `SeqQuest` program (Sandia National Labs.)[1] within the Perdew-Burke-Ernzerhof parameterization of the generalized gradient approximation (GGA) [17]. The atomic cores were replaced by norm-conserving nonlocal pseudopotentials of the Hamann-Schlüter-Chiang [18] (for C and Al) and Troullier-Martins [19] (for Au) types. The atomic-orbital-like localized basis sets were constructed using double-ζ-plus-polarization quality of Gaussian basis sets. Grid spacing of 0.3 Bohr was adopted for the real-space projection of density and to calculate the associated Hartree and exchange-correlation potentials and energy functionals. For the infinite $C_{60}$ chain models, rectangular simulation boxes with periodic boundary condition were used, and the inter-chain distance was kept to a minimum of 15 Å to avoid lateral interactions. For the metal−$C_{60}$ chain–metal junction models, again to supress the lateral inter-chain interactions, we adopted two-dimensionally periodic three-layer 6 × 6 Au(111) and Al(111) slabs (36 atoms per layer) and corrresponding simulation boxes with large area of ~250 Å$^2$/molecule (with the $C_{60}$–nearest-neighbor-cell image $C_{60}$ carbon-carbon distance of ~1.2 nm). Accordingly, only a single $\Gamma$ $\vec{k}_\parallel$-point was sampled within the surface Brillouin zone.

---

[1] Schultz PA. SeqQuest Project Albuquerque: Sandia National Laboratories (http://dft.sandia.gov/Quest).

For the MGF calculations [20] of the metal − $C_{60}$ chain − metal system, we employed our in-house software detailed elsewhere [12, 21]. With the Hamiltonian and overlap matrices H and S for junction models obtained from DFT, the retarded Green's functions at energy $E$ near the Fermi level $E_F$ were computed as

$$G(E) = (ES - H + \Sigma_1 + \Sigma_2)^{-1}, \tag{1}$$

where the self-energies, $\Sigma_{1/2}$, provide the *ab initio* broadening and shift of molecular energy levels due to the coupling with the metal electrode 1/2. To calculate the surface Green's functions, important ingredients in computing self-energies correctly, independent DFT calculations for the bulk systems that correspond to the electrode regions, i.e., one $C_{60}$ unit for the infinite $C_{60}$ chains (three-layer Au and Al slabs for the metal-$C_{60}$ chain-metal junctions), were carried out with a single $\Gamma$ $\vec{k}_\parallel$-point sampled along the surface-parallel direction and six (four) $\vec{k}_\perp$-points sampled along the surface-normal direction for the $C_{60}$ (metal slabs). The transmission function $T(E)$ were finally computed according to

$$T(E) = Tr[\Gamma_1(E)G(E)\Gamma_2(E)G^+(E)], \tag{2}$$

where $\Gamma_{1/2} = i[\Sigma_{1/2} - \Sigma_{1/2}^+]$ are the broadening matrices, with an energy scanning step of 0.01 eV at around $E_F$. For the metal−$C_{60}$ chain–metal junction models, we checked that the rather small number of metallic layers used as electrodes (three) and $\vec{k}_\parallel$-point sampling (single Γ-point), limitations im-posed by the necessity to deal with very large junction models, do not meaningfully affect the computational accuracy. The details of convergence checks are provided in Supplementary data Fig. S1 (See also [22]). The spatial and energetic distribution of energy levels and nature of the channels for the transmission were analyzed in terms of energy- and space-resolved atom projected density of states (PDOS) or local density of states (LDOS).

**3.    Results and Discussion**

3.1.    *Assessment of the intrinsic switching capacity of infinite $C_{60}$ chains*

Before examining the $C_{60}$ nanowires sandwiched between metal electrodes, we first discuss the charge transport properties of infinite unbound (U) and bound (B) $C_{60}$ chains. For the non-polymerized $C_{60}$ chains, there exist ambiguities in terms of the intermolecular distances and relative $C_{60}$-$C_{60}$ configurations. Within the room-temperature pristine face-centered-cubic (fcc) solid phase, the $C_{60}$ molecules are known to be freely rotating with the $C_{60}$–$C_{60}$ center-to-center distance of $d(C_{60}$–$C_{60}) = 10.02$ Å [23, 24]. Moreover, it appears that within the thin-film phase the fullerene assembly is additionally affected by the substrates [7, 25]. So, we have adopted in this work two literature inter-fullerene distance values, $d(C_{60}$-$C_{60}) = 10.49$ Å [7] and 10.16 Å [25], and three relative $C_{60}$–$C_{60}$ configurations, parallel hexagonal face-to-hexagonal face (H, Fig. 1a), pentagonal face-to-pentagonal face (P), and hexagon-hexagon (66) bond-to-66 bond (HH) orientations (See Supplementary data Fig. S2) [26]. We confirmed that the qualitative conclusions presented in this paper are not modified by the specifics of inter-fullerene geometries of non-polymerized chains. Below, we will mainly discuss the results obtained for $d(C_{60}$–$C_{60}) = 10.49$ Å, while the corresponding data for $d(C_{60}$–$C_{60}) = 10.16$ Å are summarized in Supplementary data Figs. S3 ~ S5.

   Compared with the unbound chain cases, the inter-fullerene geometries are more well defined for the bound $C_{60}$ chains. The polymerized $C_{60}$ chains are characterized by the [2+2] cycloaddition bond formation within the HH configuration [2, 8, 27, 28], i.e., the two 66 double bonds are broken to form four-atom carbon ring (Fig. 1b). Our DFT geometry optimizations show that the cycloaddition polymerization induces the contraction of the $C_{60}$–$C_{60}$ center-of-mass distance to $d(C_{60}–C_{60})$ = 9.08 Å with the inter-fullerene carbon–carbon distance of 1.59 Å, which is in good agreement with experimental data [2, 7, 9, 27].

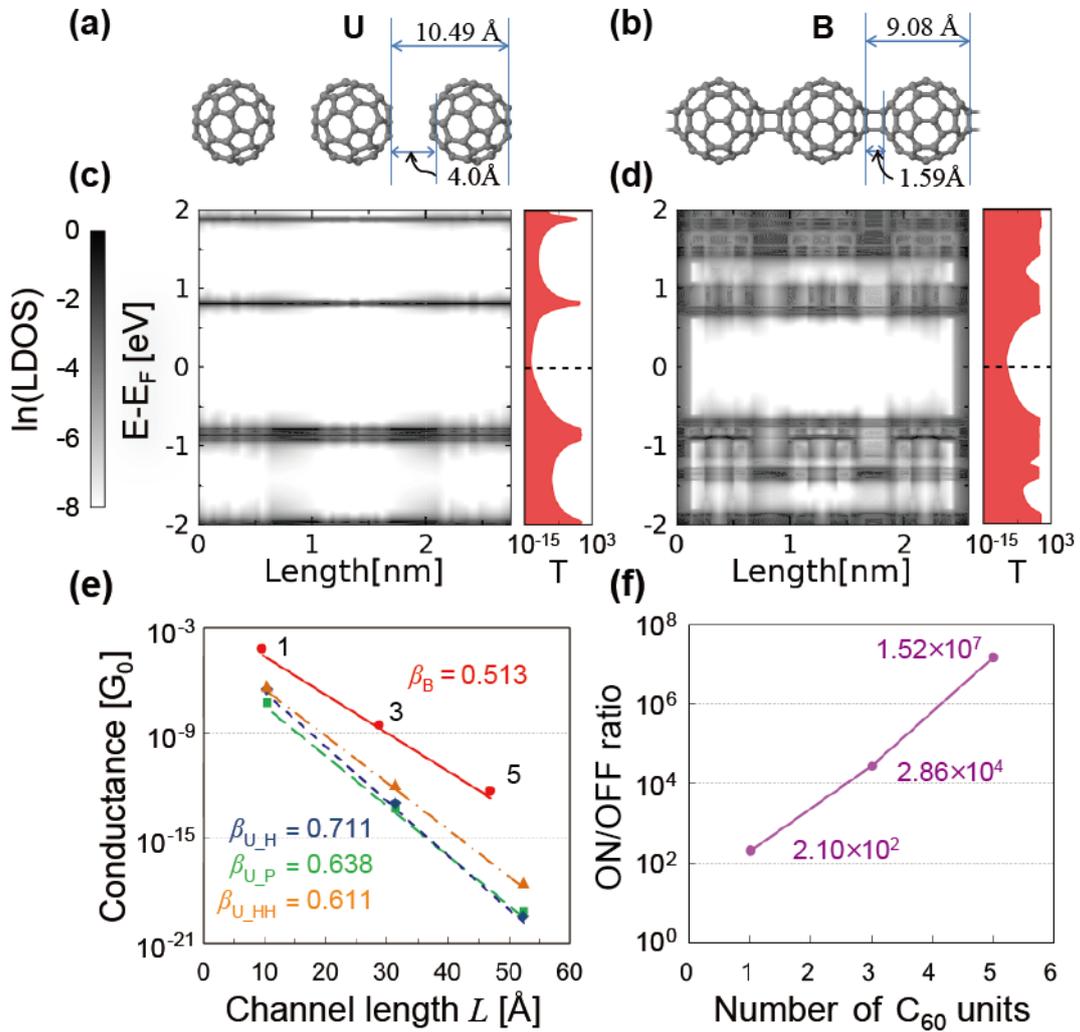

**Fig. 1** Model structures of infinite (a) H non-polymerized and (b) polymerized $C_{60}$ chains. LDOS spectra of (c) unbound and (d) bound chains calculated with an arbitrary broadening of $\delta$ = 0.05 eV

through MGF shown in log scale. The corresponding transmission spectra calculated for the three-$C_{60}$-unit scattering section with $\delta = 0.05$ eV are shown on the right-hand side. (e) The conductance length scalings of infinite $C_{60}$ chains extracted from the one-, three-, and five-$C_{60}$-unit scattering section cases. The red solid line indicates the bound chain case, while the blue dashed line, green long-dashed line, and orange dot-dashed line correspond to unbound chains with the H, P, and HH relative $C_{60}$-$C_{60}$ orientations, respectively. (f) The corresponding ON/OFF switching ratios with the unbound=OFF/bound=ON assignment.

The energy level distributions and electron transport characteristics of infinite $C_{60}$ chains were analyzed through the LDOS and transmission data calculated with an arbitrary broadening of $\delta = 0.05$ eV, as shown in Figs. 1c and 1d for the unbound and bound chains, respectively. Comparison of the two LDOS plots clearly shows that the polymerization via [2+2] cycloaddition results in strong orbital hybridizations and corresponding energy level broadening or band formation (Fig. 1d). For the non-polymerized case, on the other hand, the identity of individual $C_{60}$ is well preserved (Fig. 1c). The energy level broadening (band formation) in the bound $C_{60}$ chains is very strong in both the valence and conduction band regions, which results in a decrease of the band gap from 1.68 eV of the unbound chain to 1.34 eV of the bound chain. Note that, although the calculated highest occupied molecular orbital (HOMO) –lowest unoccupied molecular orbital (LUMO) gaps are comparable to those extracted from photoemission experiments for negative gas phase clusters [29], they should have been underestimated compared with the HOMO–LUMO gaps seen by photoemission or tunneling [30-32] due to the well-known self-interaction errors within GGA [33, 34]. This point will be further discussed later. The detailed nature of hybridized frontier molecular orbitals in polymerized chains that participate in the charge transport can be understood by visualizing them in real space, and in [11], taking the trimer case as a representative example, we demonstrated that the orbitals intrinsically delocalized throughout the chain or the localized molecular orbitals with many energetically close neighboring orbitals become the main charge transport pathways. The localized/delocalized nature of energy levels in the unbound/bound chains are well reflected on the corresponding transmission spectrum calculated for the three-unit $C_{60}$ section, as shown on the right-hand side of Figs. 1c and 1d, from which we find that the polymerization enhances the zero-bias transmission by several orders of magnitudes.

The spatially decaying (rather than propagating) complex Bloch $k$ states within the HOMO-LUMO gaps of $C_{60}$ wires can be further characterized by the scaling of conductance with wire length $L$,

$$G = G_0 \exp(-\beta L), \tag{3}$$

where $G_0 = 2e^2/h = 77.6$ μS, or the decay rate $\beta = 2\, \text{Im}(k)$. The tunneling decay factor $\beta$, which has been traditionally extracted from the complex band structure [35], is a function of energy, and here we will focus on the Fermi level or midgap point that approximately corresponds to the branch point of complex band structure. Data obtained for the unbound and bound infinite chains, computed by taking one, three, and five $C_{60}$ units as the scattering region, are shown in Fig. 1e. In agreement with their electronic structures, polymerized $C_{60}$s have consistently higher conductance values ("ON" state) compared with their non-polymerized counterparts in the HH, H, and P configurations ("OFF" state). The enhancement of intrinsic conductance of $C_{60}$ chains with polymerization can be quantified by the decay rate of the polymerized chain, $\beta_B = 0.513$, which is much smaller than that of its non-polymerized counterparts, $\beta_{U\_H} = 0.711$, $\beta_{U\_P} = 0.638$, and $\beta_{U\_HH} = 0.611$ for the H, P, and HH configurations, respectively. Note that the difference in $\beta_B$ and $\beta_U$ results in the increase of ON/OFF ratio with the chain length $L$ (Fig. 1f), but at the cost of a decrease in the conductance magnitude. Thus, it can be expected that the choice of an optimal chain length in the practical device realization will have to be determined by the compromise of the two factors. In the following studies of junction models, due to the negligible differences among the intrinsic properties of three unbound chains with respect to those of the bound chain counterpart, we will adopt only the H unbound configuration (Fig. 1a).

Before moving on, we comment on the approach we adopted here to extract the $\beta$ values and switching ratios of hypothetical infinite chains. Note that this novel way of extracting $\beta$ within the MGF formalism is equivalent to the complex band structure approach for infinite chains [35], but has the added benefit of enabling us to estimate the switching ratios of a finite section within the infinite chain. First, we emphasize that the finite conductance values within the HOMO−LUMO gaps (Figs. 1c and 1d) result from the imaginary electrons crossing virtual gap states (i.e., there exists no electron reservoir that can supply real incident electrons) with an arbitrary impulse energy defined by the broadening factor $\delta = 0.05$ eV. These virtual gap states correspond to the complex band structure and originate the metal-induced gap states (MIGS) in the real metal−seminconductor interfaces. The choice of the broadening factor $\delta$, which replaces the broadening matrix $\Gamma$ (defined in Sec. 2), is completely arbitrary. However, the data obtained with different $\delta = 0.05, 0.01, 0.005$, and $0.001$ eV summarized in Fig. 2 (see also Supplementary data Table S1) demonstrate that, irrespective of the fact that the absolute conductance values are rather meaningless, our computational procedure provides numerically robust (essentially identical) decay parameters and ON/OFF ratios. We finally note that our scheme can be more generally applied to the "interface-induced gap states" in semiconductor heterostructures [36] where semiconductor slabs should be considered as "electrodes", and in [37] we have successfully employed the method to characterize the tunneling currents in Si−SiO$_2$−Si junctions.

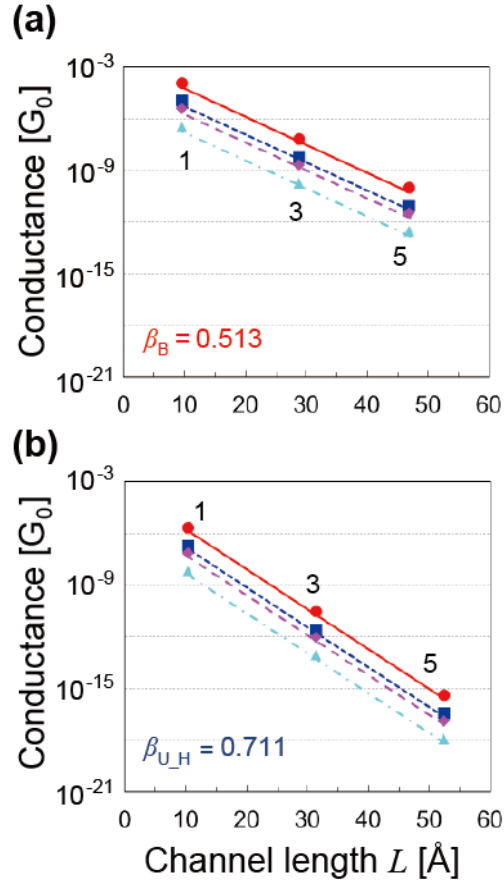

**Fig. 2** The conductance length scalings of infinite $C_{60}$ chains calculated for the one-, three-, and five-$C_{60}$-unit scattering sections for the (a) bound and (b) unbound chain cases with different broadening factors. Red solid, blue dotted, magenta dashed, and cyan dot-dashed lines represent $\delta$ = 0.05, 0.01, 0.005, and 0.001 eV cases, respectively. Different $\delta$ values result in identical ON/OFF ratios of $2.10 \times 10^2$, $2.86 \times 10^4$, and $1.52 \times 10^7$, for the monomer, dimer, and pentamer cases, respectively.

3.2. *Analysis of the switching characteristics of STM junctions*

To analyze the working principles of switching behavior realized through STM experiments [9, 10], we next constructed junction models by placing $C_{60}$ chains between Au(111) and Al(111) slabs with asymmetric left electrode (substrate)-(bottom) $C_{60}$ distance $d_1$ and right electrode (STM tip)-(top) $C_{60}$ distance $d_2$, as shown in Figs. 3a and 3b for the unbound (U) and bound (B) chain cases, respectively.

Simulating the constant-height STM mode, we fixed the electrode-electrode gap distance $l$.

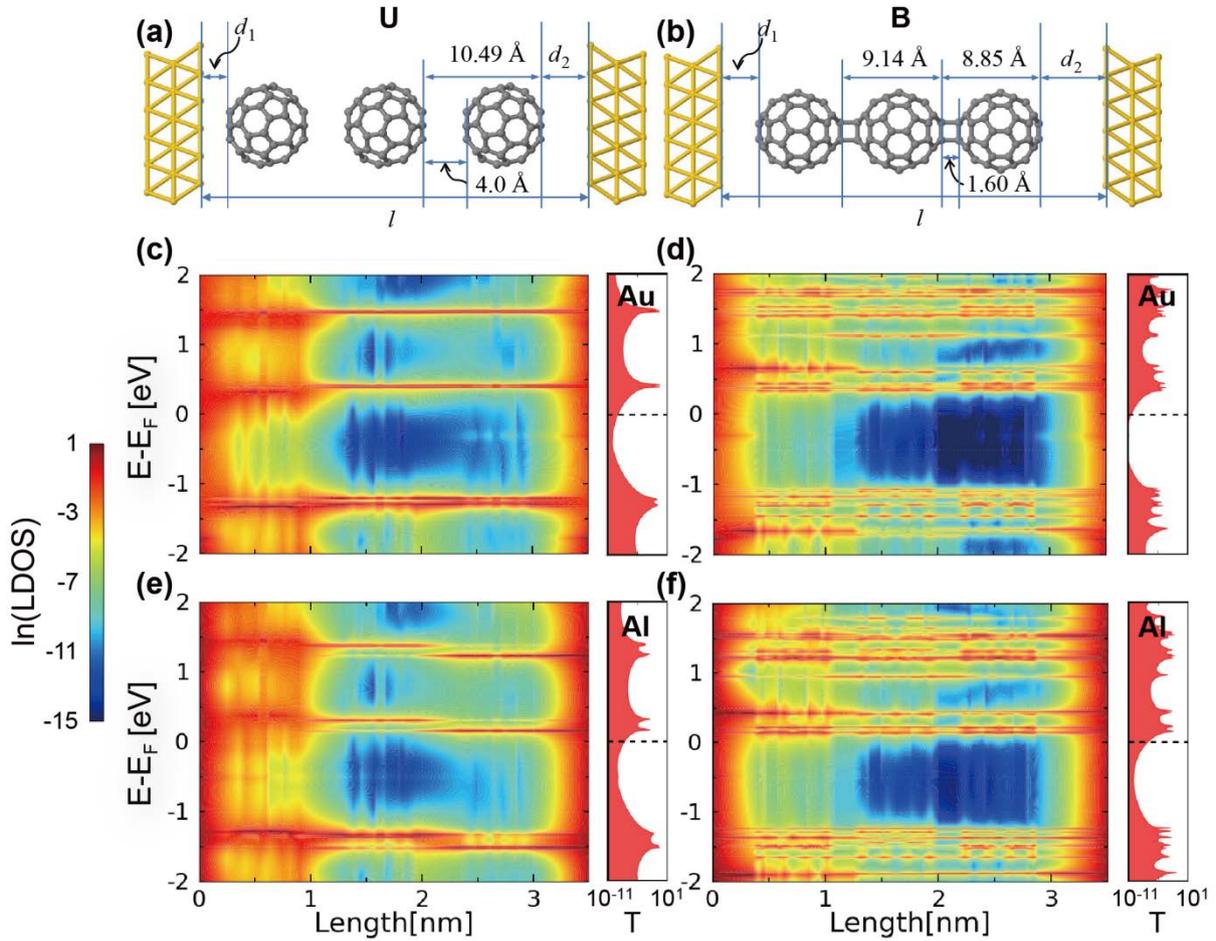

**Fig. 3** Side views of the $C_{60}$ trimer-STM junction models with a fixed electrode-electrode gap distance of $l$ = 34.98 Å: (a) non-oligomerized chain with $d_1$ = 2.5 Å and $d_2$ = 5.0 Å and its (b) oligomerized counterpart with $d_1$ = 3.61 Å and $d_2$ = 6.11 Å. The corresponding LDOS and transmission functions (shown in log scale) for the (c, d) Au and (e, f) Al electrode cases.

In modeling the junctions, we have introduced several idealizations. First, $C_{60}$s assume fcc close-packing arrangements within the well-ordered thin film phase [7, 9, 10], which leads to the tilting of linear $C_{60}$ chains from the surface-normal direction by 35.26°. We here simplified the situation by considering an isolated upright $C_{60}$ chain. In addition, whereas random $C_{60}$ orientations would be more realistic for non-oligomerizaed $C_{60}$ chain models [23, 24], we adopted the H inter-fullerene configurations for the straightforward comparison with infinite non-polymerized $C_{60}$ chain cases (Sec. 3.1). In terms of electrode models, whereas the experiments were carried out with the Si(111) $\sqrt{3}\times\sqrt{3}R30°$-Ag substrate and the electrochemically etched Pt-20%Ir STM tip, we adopted the simpler Al-Al and Au-Au electrode configurations and assessed the effects of metal work function differences. Next, experimental Au–$C_{60}$ and Al–$C_{60}$ distances are currently unavailable and theoretical values sensitively depend on the choice of the DFT exchange-correlation functional due to the van der Waals interaction character [38]. We thus followed our earlier approach of considering several plausible contact distances and focusing on the electronic and transport characteristics that are robust with respect to the variation of metal–$C_{60}$ distances [11]. More specifically, we will discuss in the main text below the data obtained with $d_1 = 2.5$ Å for the unbound chain case. The corresponding data from $d_1 = 3.0$ Å (unbound chain cases) are provided in Supplementary data (Fig. S6 and Table S2), which show that our results are qualitatively unaffected by the variation of $d_1$. Finally, to model the STM junctions in the constant-height mode, the distance between $C_{60}$ and top metal electrode (that represent the STM tip) $d_2$ was fixed at 5.0 Å, a typical value for the STM experiments [39, 40], for the non-oligomerized case.

Based on the reference junction models based on unbound $C_{60}$ chains, we constructed oligomerized chain counterparts by maintaining the center of gravity, as suggested in the experiments [9, 10]. This oligomerization induces a 2.22 Å of contraction of the $C_{60}$ trimer chain length (from 27.46 Å for the unbound chain to 25.24 Å for the bound chain), which is consistent with the experimental estimation [9, 10]. So, for the $d_1 = 2.5$ Å and $d_2 = 5.0$ Å unbound trimer case with the constant electrode–electrode gap distance $l = 34.98$ Å, the contact distances of the corresponding bound chain junction model amount to $d_1 = 3.61$ Å and $d_2 = 6.11$ Å (Table 1). The contraction of $C_{60}$ chain length unpon oligomerization increases with the number of $C_{60}$ units, so a similar choice of the contact distance for the reference unbound chains results in corresponding increases of $d_1$ and $d_2$ for longer bound chains: e.g., the unbound pentamer junction with $d_1 = 2.5$ Å and $d_2 = 5.0$ Å ($l = 43.56$ Å) results in $d_1 = 4.96$ Å and $d_2 = 7.46$ Å for the bound counterpart.

| Metal | Configuration | Contact distance [Å] | | Conductance [$G_0$] | ON/OFF ratio |
|---|---|---|---|---|---|
| | | $d_1$ | $d_2$ | | |
| Au | Unbound | 2.5 | 5.0 | $1.70 \times 10^{-9}$ (ON) | 28.64 |
| | Bound | 3.61 | 6.11 | $5.93 \times 10^{-11}$ (OFF) | |
| Al | Unbound | 2.5 | 5.0 | $3.83 \times 10^{-8}$ (OFF) | 3.91 |
| | Bound | 3.61 | 6.11 | $1.50 \times 10^{-7}$ (ON) | |

**Table 1** Metal-$C_{60}$ contact distances $d_1$ and $d_2$, conductance values, and ON/OFF ratios of the $C_{60}$ trimer STM switch models for the fixed electrode-electrode distance $l = 34.98$Å.

We now discuss the electronic structure and charge transport characteristics of STM junctions using the LDOS and transmission data presented in Figs. 3c-f for the trimer cases ($d_1 = 2.5$ Å for the unbound chain). Focusing on the unbound $C_{60}$ cases, the first notable feature common to both Au (Fig. 3c) and Al (Fig. 3e) electrodes is the significant MIGS [36] or the hybridization between metal surface bands and molecular orbitals at the left electrode (substrate)–(bottom) $C_{60}$ interfaces. The MIGS are dominated by the $C_{60}$ LUMOs, which agrees with the conventional characterization of fullerenes as *n*-channel materials. Moreover, comparing the Au and Al cases, it is evident that the strong charge transfer at the left Au-$C_{60}$ and Al-$C_{60}$ interfaces leads to the Fermi-level pinning such that the LUMO is located at ~0.35 eV above $E_F$. As discussed earlier, this value should have been underestimated compared with experimental data [30-32, 34]. However, since the STM experiment probes the energy range right below the LUMO and we focus on differences (between oligomerized and deoligomerized chains or Au and Al electrodes) rather than absolute values, our conclusions will not be qualitatively modified. We previously pointed out that, due to its strong electron accepting capacity, the pinning energy position for $C_{60}$ should be rather insensitive for typical contact distances of about 2 – 3 Å between substrate metals and $C_{60}$s deposited on them [11].

Contrasted with the left metal (substrate) −$C_{60}$ interfaces, the much longer contact distance ($d_2 = 5$ Å) for the right (top) $C_{60}$-STM tip interfaces results in greatly reduced charge transfer and level broadening or significantly reduced Fermi-level pinning and enhanced vacuum level alignment. Because of the differences in work functions of Au, $C_{60}$, and Al (5.1, 4.7, and 4.3 eV, respectively), the energy levels of the rightmost $C_{60}$ in the Au and Al electrode cases shift in the opposite directions relative to the pinned leftmost $C_{60}$ levels. This can be more clearly seen in the PDOS of the leftmost (bottom) and rightmost (top) $C_{60}$s shown in Fig. 4a: the LUMOs shift upward from 0.32 to 0.41 eV for the Au case but downward from 0.37 to 0.15 eV for the Al counterpart.

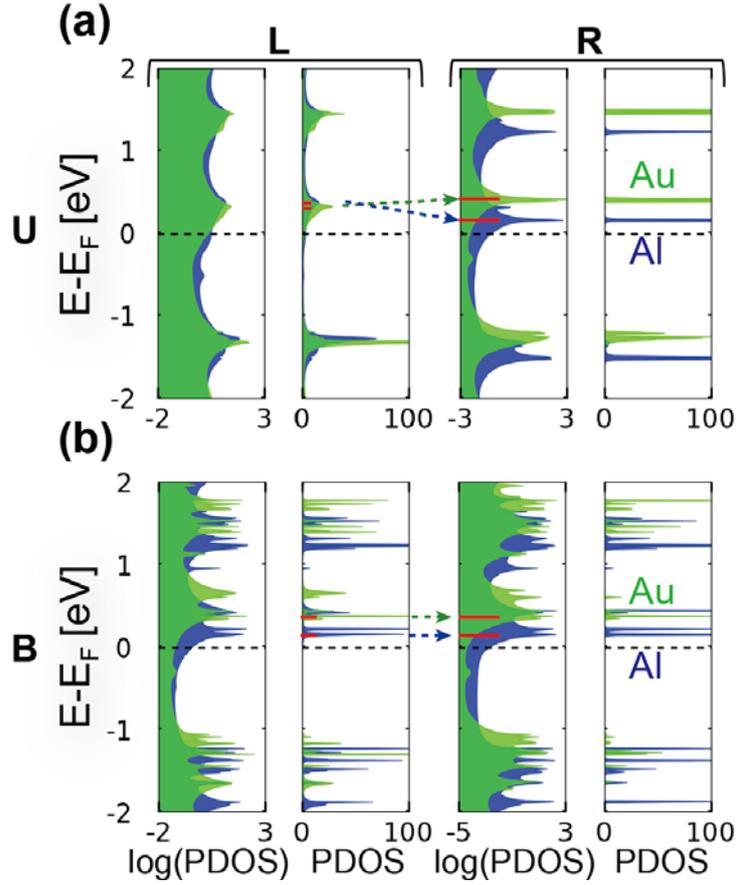

**Fig. 4** PDOS plots in log and normal scales of the leftmost (L) and rightmost (R) $C_{60}$ unit within the (a) non-oligomerized and (b) oligomerized chains inside the $C_{60}$ trimer-STM junction models. For PDOS plots, L and R respectively indicate the leftmost and rightmost $C_{60}$ in contact with Au (green) or Al (blue) both in log (left) and normal (right) scale. The red lines indicate the location of LUMO, and the dotted arrows represent the LUMO level shifts.

As characterized for the infinite chain cases, the oligomerization of $C_{60}$ chains results in the delocalization of energy levels throughout all three $C_{60}$ molecules. Due to the increase of substrate-bottom $C_{60}$ distance $d_1$, these "bands" collectively shift upward/downward for the Au/Al electrode case. This feature can be again more clearly seen in the PDOS of the bottom and top $C_{60}$s shown in

Fig. 4b.

Transmission curves are presented on the right-hand side of the LDOS spectra in Figs. 3c-f (see also Table 1). The behaviors of energy level (especially LUMO) shifts are directly reflected on the junction charge transport properties. The collective upward/downward shift of energy levels upon $C_{60}$ chain oligomerization for the Au/Al electrodes cause the LUMOs, which dominate the conductance at $E_F$, to move away from/toward $E_F$. Another factor that affects the conductance magnitude is the increase of electrode–$C_{60}$ distances $d_1$ and $d_2$ for the bound chains, which should reduce the magnitude of the tunneling current. The upward LUMO shift in the $C_{60}$–STM tip region and the increase of $d_1$ and $d_2$ upon chain oligomerization cooperatively leads to the assignment of unbound = ON/bound = OFF switching states for the Au electrode cases and a switching ratio of 28.64. On the other hand, the competition of the two effects (the downward LUMO shift in the $C_{60}$-STM tip region and the increase of $d_1$ and $d_2$ upon chain oligomerization) results in the opposite unbound = OFF/bound = ON assignment for the Al electrode case and a much smaller switching ratio of 3.91.

To further characterize the general nature of charge transport properties of STM junctions, we again considered the conductance scaling with channel length $l$. The results are summarized in Figs. 5a and 5b for the Au and Al electrode cases, respectively, and the resulting ON/OFF switching ratios for different $C_{60}$ units are presented in Fig. 5c. The conductance scalings as well as switching ratios show critical dependence on the electrode metal species: for the Au electrode case, unbound $C_{60}$ chains show higher conductance magnitudes and lower decay rate ($\beta_U = 0.662$) than their bound counterparts ($\beta_B = 0.838$), which is opposite to the behavior of infinite chains (Fig. 1e). For the Al electrode case, due to the above-described opposite effects of the energy level shift in the $C_{60}$–STM tip gap region and its distance variation, we find only negligible differences between unbound and bound chain cases. We thus conclude that the intrinsic switching capacity of infinite $C_{60}$ chains is not properly manifested in the constant-height mode STM device configuration.

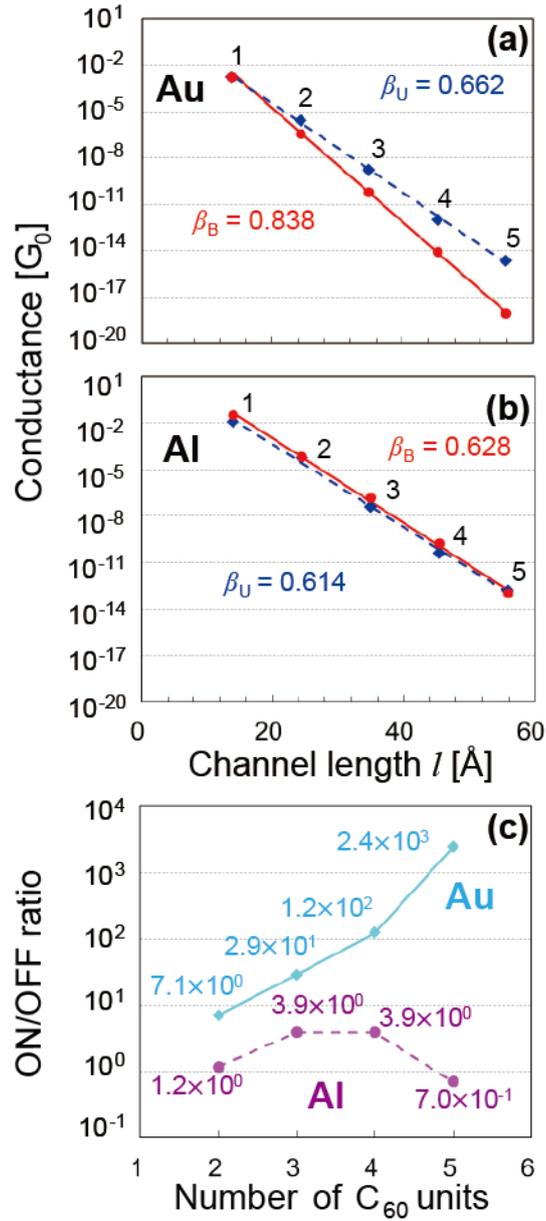

**Fig. 5** The conductance length scalings of the STM junction models based on one- to five-unit $C_{60}$ chains sandwiched between (a) Au and (b) Al electrodes. The red solid and blue dashed lines indicate the bound and unbound chain cases, respectively. (c) The corresponding ON/OFF switching ratios for the Au (cyan solid line) and Al (magenta dashed line) electrode junctions. The former case was derived from the unbound = ON/bound = OFF assignment, while the latter was obtained with the unbound = OFF/bound = ON assignment.

## 3.3. *Proposal of a nanoelectromechanical device architecture with an ultra-high ON/OFF switching ratio*

So far, we have shown that, in agreement with experimental conclusions, the intrinsic change of electronic property accompanying the $C_{60}$ chain oligomerization does not manifest within the constant-height STM junction architecture. Remind that our previous analysis of ideal infinite $C_{60}$ chains provided the formal ground to devise a device architecture that achieves an ultra-high switching ratio. This naturally suggests us to consider an approach to fully utilize the intrinsic switching capacity of the $C_{60}$ chain oligomerization/deoligomerization in the device settings and provide electrode metal-independent device characteristics while maintaining an enhanced ON/OFF switching ratio. We now demonstrate that adopting a nanoelectromechanical (NEM) device architecture that simply takes the minimized contact distances for both ends of the $C_{60}$ chain can achieve the goal. Such mechanically pressed contacts could be established with, e.g., C-AFM that has demonstrated atomic resolution [16]. The NEM device models based on $C_{60}$ trimers are shown in Figs. 6a and 6b for the unbound ($d_1 = d_2 \equiv d = 2.5$ Å and $l = 32.49$ Å) and bound cases ($d = 2.5$ Å and $l = 30.27$ Å), respectively (See also Table 2).

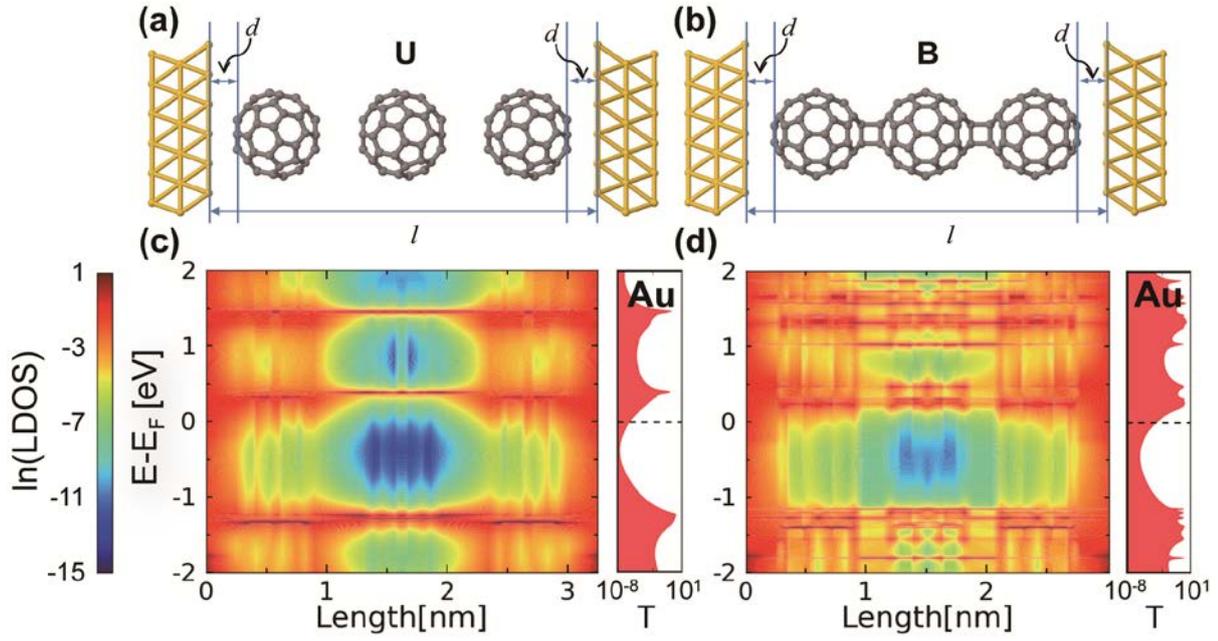

**Fig. 6** Side views of the $C_{60}$ unbound and bound trimer-NEM junction models with the fixed electrode-$C_{60}$ distance $d_1 = d_2 \equiv d = 2.5$ Å: (a) the non-oligomerized case with the electrode-electrode gap distance $l = 32.49$ Å and (b) the oligomerized counterpart with $l = 30.27$ Å. The corresponding (c) unbound- and (d) bound-chain LDOS and transmission functions (shown in log scale) for the Au electrode case.

| Metal | Configuration | Gap distance $l$ [Å] | Contact distance $d$ [Å] | Conductance [$G_0$] | ON/OFF ratio |
|---|---|---|---|---|---|
| Au | Unbound | 32.49 | 2.5 | $6.00 \times 10^{-7}$ (OFF) | 301.5 |
| | Bound | 30.27 | 2.5 | $1.81 \times 10^{-4}$ (ON) | |
| Al | Unbound | 32.49 | 2.5 | $9.38 \times 10^{-7}$ (OFF) | 1034.2 |
| | Bound | 30.27 | 2.5 | $9.70 \times 10^{-4}$ (ON) | |

**Table 2** Electrode-electrode gap distances $l$, metal-$C_{60}$ contact distances $d$, conductance values, and ON/OFF ratios of the $C_{60}$ trimer NEM switch models.

The LDOS and transmission spectra for the Au electrode-based junction are presented in Figs. 6c and 6d for the unbound and bound chain cases, respectively. Since the metal-dependent energy level shift is suppressed due to the Fermi-level pinning in both force-feedbacked contacts in the NEM model [11], we obtain similar results for the Au and Al electrodes and here show the data for the Au electrode case only. Comparing with Fig. 3, the symmetric and similar energy level distributions including Fermi-level pinning at both contacts in the unbound as well as bound chain cases clearly allow the intrinsic (bulk) chain electronic structures to become the major discriminating factor of the two junctions. Therefore, the comparison of transmission functions of unbound and bound trimers (located on the right-hand side of Figs. 6c and 6d) show the effects of "band" formation in the bound chain case, which is similar to the situation in infinite $C_{60}$ wires (Figs. 1c and 1d). Unlike in the constant-height STM switches, but as could be expected from the infinite chains, we consistently obtain unbound = OFF/bound = ON switching states for both Au and Al electrodes and the switching ratios are also orders of magnitudes larger than those from constant height STM switches (301.5 and 1034.2 for the Au and Al electrodes, respectively) (Table 2).

Whether or not the intrinsic charge transport properties and associated switching capacity of infinite $C_{60}$ chains are properly transferred to the NEM switches can be determined by the device conductance scaling with fullerene channel length $l$, as shown in Figs. 7a and 7b. The corresponding ON/OFF switching ratios of the NEM switches are summarized in Fig. 7c. We observe that, unlike with the constant height STM switches (Figs. 5a and 5b), the decay rates are almost independent of the electrode metal species ($\beta_B$ = 0.403 and 0.410 for the Au and Al electrodes, respectively, and $\beta_U$ = 0.644 and 0.656 for the Au and Al electrodes, respectively). While these values are slightly smaller than those of infinite $C_{60}$ chains (Fig. 1e), the overall decaying characteristics of infinite (un)bound $C_{60}$ chains are preserved within the NEM junctions. Also, we find that the switching ratio of the NEM devices would be about only an order of magnitude smaller than that of the infinite chains. So, we obtain the metal-independent unbound = OFF/bound = ON switching characteristics as well as very large switching ratios that are several orders of magnitude larger than those of constant height STM switches. While these impressive switching ratios could be reduced in more realistic experimental situations due to, e.g., tilting of $C_{60}$ chains, we still expect a switching performance that is drastically improved over that of the experimentally-realized constant-height mode STM switches.

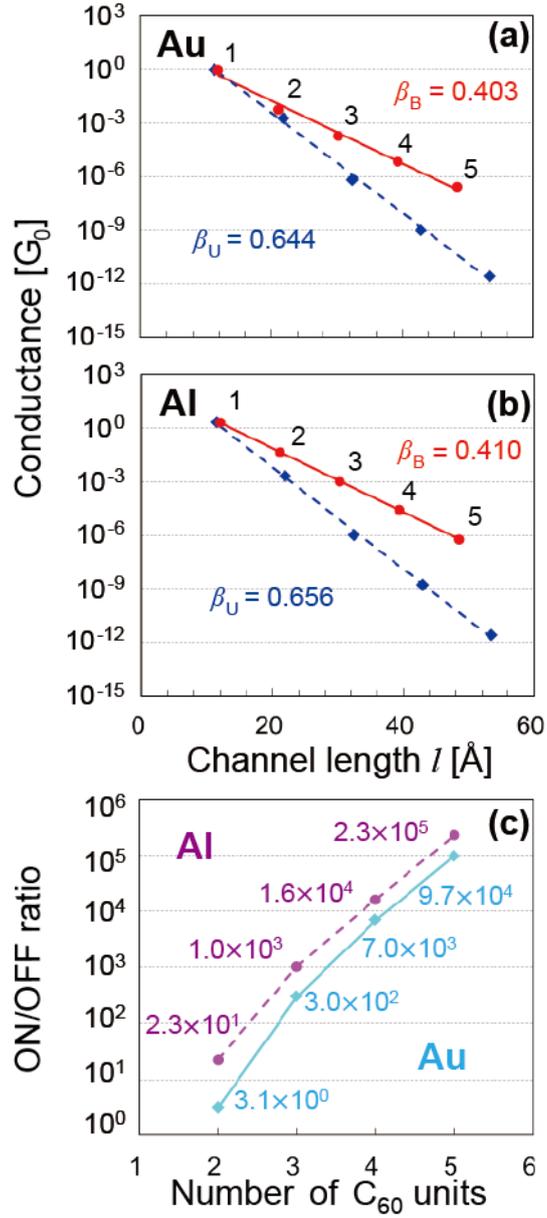

**Fig. 7** The conductance length scalings of the NEM junction models derived from one-to-five- unit $C_{60}$ chains for the (a) Au and (b) Al electrode junctions. The red solid and blue dashed lines indicate the bound and unbound chain cases, respectively. (c) The corresponding ON/OFF switching ratios for the Au (cyan solid line) and Al (magenta dashed line) electrode junctions.

**4. Conclusions**

In summary, in view of recent experimental realization of molecular-scale switches based on the reversible formation and breaking of linear $C_{60}$ chains using STM, we have carried out comparative first-principles computational studies of bound and unbound $C_{60}$ chains sandwiched between Al(111) and Au(111) electrodes within the asymmetric STM junction geometry and proposed a new symmetric NEM memory device architecture that can achieve drastically larger ON/OFF switching ratios. Considering infinite unbound and bound $C_{60}$ chains, we first showed that oligomerized $C_{60}$ chains with much stronger orbital hybridizations should in principle achieve several orders of magnitudes higher conductance values compared with their non-ologomerized counterparts. However, we found that the large intrinsic switching capacity of $C_{60}$ wires is difficult to realize within the STM junction setting: Because of the dominant role of large $C_{60}$–STM tip distances, the ON/OFF switching ratio was found to be relatively small. Moreover, the assignment of ON/OFF switching states was predicted to sensitively depend on electrode metal species (unbound = ON/bound = OFF for Au and the opposite for Al) because the energy level bending from the pinned energetic position at the substrate-side contact toward the STM tip-side vacuum gap can change directions according to the substrate electrode metal work functions. Based on the analysis of ideal infinite $C_{60}$ chains that hinted the possibility of achieving a much improved device performance, we finally seek a method to fully exploit the intrinsic switching capacity of $C_{60}$ wires and minimize the effects of metal electrodes. We proposed a device architecture where the $C_{60}$–metal electrode contact distances are maintained in an NEM fashion and demonstrated that they can indeed achieve ultra-high ON/OFF switching ratios with metal-independent unbound = OFF/bound = ON switching states. We expect such an NEM device could be realized with, e.g., C-AFM. The intrinsic difference between the constant-height STM and NEM architectures was substantiated by comparing the conductance length scalings of the two junctions with that of infinite $C_{60}$ chains, which is, to the best of our knowledge, a novel method to characterize molecular-scale switches. So far, only a few re-ports have been made on the first-principles quantum charge transport calculations of $C_{60}$ chains [11, 41, 42], and, to our knowledge, the present work represents the most systematic large-scale study that has considered various parameters such as the chain oligomerization, electrode metal species, metal-fullerene contact distance, and inter-fullerene distance. Our findings on the critical role of the interplay between the energy level pinning and depinning at contacts and the changes of intrinsic electronic structures of $C_{60}$ chains in determining junction characterisitics should be relelvant for the future development of molecular devices.

**Acknowledgments**

This work was supported by Global Frontier R&D Program (No. 2013-073298), Basic Science Research Grant (No. 2012R1A1A2044793), Nano·Material Technology Development Program (2012M3A7B4049888), and Global Ph.D. Fellowship Program of the National Research Foundation funded by the Ministry of Science, ICT & Future Planning of Korea. Computational resources were provided by the KISTI Supercomputing Center (KSC-2012-C2-20).

**References**

[1]     van der Molen SJ, Liljeroth P. Charge transport through molecular switches. J Phys Condens Matter 2010;22(13):133001.

[2]     Rao AM, Zhou P, Wang KA, Hager GT, Holden JM, Wang Y, et al. Photoinduced polymerization of solid $C_{60}$ films. Science 1993;259(5097):955-7.

[3]     Iwasa Y, Arima T, Fleming RM, Siegrist T, Zhou O, Haddon RC, et al. New phases of $C_{60}$ synthesized at high-pressure. Science 1994;264(5165):1570-2.

[4]     Nunez-Regueiro M, Marques L, Hodeau J-L, Bethoux O, Perroux M. Polymerized fullerite structures. Phys Rev Lett 1995;74(2):278-81.

[5]     Alvarez-Zauco E, Sobral H, Basiuk EV, Saniger-Blesa JM, Villagran-Muniz M. Polymerization of $C_{60}$ fullerene thin films by UV pulsed laser irradiation. Appl Surf Sci 2005;248(1-4):243-7.

[6]     Moret R, Wagberg T, Sundqvist B. Influence of the pressure-temperature treatment on the polymerization of $C_{60}$ single crystals at 2GPa-700K. Carbon 2005;43(4):709-16.

[7]     Hassanien A, Gasperic J, Demsar J, Musevic I, Mihailovic D. Atomic force microscope study of photo-polymerized and photo-dimerized epitaxial $C_{60}$ films. Appl Phys Lett 1997;70(4):417-9.

[8]     Wang GW, Komatsu K, Murata Y, Shiro M. Synthesis and X-ray structure of dumb-bell-shaped $C_{120}$. Nature 1997;387(6633):583-6.

[9]     Nakaya M, Tsukamoto S, Kuwahara Y, Aono M, Nakayama T. Molecular scale control of unbound and bound $C_{60}$ for topochemical ultradense data storage in an ultrathin $C_{60}$ film. Adv Mater 2010;22(14):1622-5.

[10]    Nakaya M, Aono M, Nakayama T. Molecular-scale size tuning of covalently bound assembly of $C_{60}$ molecules. ACS Nano 2011;5(10):7830-7.

[11]     Lee GI, Kang JK, Kim YH. Metal-independent coherent electron tunneling through polymerized fullerene chains. J Phys Chem C 2008;112(17):7029-35.

[12]     Kim YH, Jang SS, Jang YH, Goddard WA. First-principles study of the switching mechanism of [2]catenane molecular electronic devices. Phys Rev Lett 2005;94(15):156801.

[13]     Kim YH. Toward numerically accurate first-principles calculations of nano-device charge transport characteristics: The case of alkane single-molecule junctions. J Korean Phys Soc 2008;52(4):1181-6.

[14]     Kim YH. Electrical and mechanical switching in a realistic [2]rotaxane device model. J Nanosci Nanotechnol 2008;8(9):4593-7.

[15]     Loh OY, Espinosa HD. Nanoelectromechanical contact switches. Nat Nano 2012;7(5):283-95.

[16]     Hong SS, Cha JJ, Cui Y. One nanometer resolution electrical probe via atomic metal filament formation. Nano Lett 2011;11(1):231-5.

[17]     Perdew JP, Burke K, Ernzerhof M. Generalized gradient approximation made simple. Phys Rev Lett 1996;77(18):3865-8.

[18]     Hamann DR, Schlüter M, Chiang C. Norm-conserving pseudopotentials. Phys Rev Lett 1979;43(20):1494-7.

[19]     Troullier N, Martins JL. Efficient pseudopotentials for plane-wave calculations. Phys Rev B 1991;43(3):1993-2006.

[20]     Datta S. Quantum transport: Atom to transistor. Cambridge, UK: Cambridge University Press; 2005.

[21]     Kim YH, Tahir-Kheli J, Schultz PA, Goddard WA. First-principles approach to the charge-transport characteristics of monolayer molecular-electronics devices: Application to hexanedithiolate devices. Phys Rev B 2006;73(23):235419.

[22]     Kim YH, Byun YM. Diameter dependence of charge transport across carbon nanotube-metal contacts from first principles. J Korean Phys Soc 2009;55(1):299-303.

[23]     Heiney P, Fischer J, McGhie A, Romanow W, Denenstein A, McCauley Jr J, et al. Orientational ordering transition in solid $C_{60}$. Phys Rev Lett 1991;66(22):2911-4.

[24]     Makarova TL. Electrical and optical properties of pristine and polymerized fullerenes. Semiconductors 2001;35(3):243-78.

[25]     Nakayama T, Onoe J, Takeuchi K, Aono M. Weakly bound and strained $C_{60}$ monolayer on the Si(111) √3×√3R30°-Ag substrate surface. Phys Rev B 1999;59(19):12627-31.

[26]     Tournus F, Charlier J-C, Melinon P. Mutual orientation of two $C_{60}$ molecules: An *ab initio* study. J Chem Phys 2005;122(9):094315.

[27]     Stephens PW, Bortel G, Faigel G, Tegze M, Janossy A, Pekker S, et al. Polymeric fullerene chains in $RbC_{60}$ and $KC_{60}$. Nature 1994;370(6491):636-9.

[28]     Adams GB, Page JB, Sankey OF, Okeeffe M. Polymerized $C_{60}$ studied by first-principles molecular-dynamics. Phys Rev B 1994;50(23):17471-9.

[29]     Yang SH, Pettiette CL, Conceicao J, Cheshnovsky O, Smalley RE. UPS of buckminsterfullerene and other large clusters of carbon. Chem Phys Lett 1987;139(3–4):233-8.

[30]     Lu XH, Grobis M, Khoo KH, Louie SG, Crommie MF. Charge transfer and screening in individual $C_{60}$ molecules on metal substrates: A scanning tunneling spectroscopy and theoretical study. Phys Rev B 2004;70(11):115418.

[31]     Kang SJ, Yi Y, Kim CY, Cho SW, Noh M, Jeong K, et al. Energy level diagrams of $C_{60}$/pentacene/Au and pentacene/$C_{60}$/Au. Synthetic Met 2006;156(1):32-7.

[32]     Schull G, Neel N, Becker M, Kroger J, Berndt R. Spatially resolved conductance of oriented $C_{60}$. New J Phys 2008;10(6):065012.

[33]     Cappellini G, Casula F, Yang JL, Bechstedt F. Quasiparticle energies in clusters determined

via total-energy differences: Application to $C_{60}$ and $Na_4$. Phys Rev B 1997;56(7):3628-31.

[34] Sau JD, Neaton JB, Choi HJ, Louie SG, Cohen ML. Electronic energy levels of weakly coupled nanostructures: $C_{60}$-metal interfaces. Phys Rev Lett 2008;101(2):026804.

[35] Tomfohr JK, Sankey OF. Complex band structure, decay lengths, and Fermi level alignment in simple molecular electronic systems. Phys Rev B 2002;65(24):245105.

[36] Mönch W. Electronic properties of semiconductor interfaces. Berlin Heidelberg: Springer; 2004.

[37] Kang J, Kim YH, Bang J, Chang KJ. Direct and defect-assisted electron tunneling through ultrathin $SiO_2$ layers from first principles. Phys Rev B 2008;77(19):195321.

[38] Hamada I, Tsukada M. Adsorption of $C_{60}$ on Au(111) revisited: A van der Waals density functional study. Phys Rev B 2011;83(24):245437.

[39] Ren H, Yang J, Luo Y. Identifying configuration and orientation of adsorbed molecules by inelastic electron tunneling spectra. J Chem Phys 2010;133(6):064702-4.

[40] Oka H, Tao K, Wedekind S, Rodary G, Stepanyuk VS, Sander D, et al. Spatially modulated tunnel magnetoresistance on the nanoscale. Phys Rev Lett 2011;107(18):187201.

[41] Ono T, Hirose K. First-principles study of electron-conduction properties of $C_{60}$ bridges. Phys Rev Lett 2007;98(2):026804.

[42] Schull G, Frederiksen T, Brandbyge M, Berndt R. Passing current through touching molecules. Phys Rev Lett 2009;103(20):206803.